\documentclass[showpacs,twocolumn,bibnotes,aps,prl]{revtex4}
\usepackage{graphicx}
\usepackage{dcolumn}
\usepackage{amstext}
\usepackage{amsmath}
\usepackage{amssymb}
\usepackage{latexsym}
\usepackage{bm}
\usepackage{amsfonts}

\begin{document}

\title{High-Q exterior whispering gallery modes in a metal-coated microresonator}
\author{Yun-Feng Xiao$^1$}
\email{yfxiao@pku.edu.cn}
\author{Chang-Ling Zou$^2$}
\author{Bei-Bei Li$^1$}
\author{Yan Li$^1$}
\author{Chun-Hua Dong$^2$}
\author{Zheng-Fu Han$^2$}
\author{Qihuang Gong$^1$}
\email{qhgong@pku.edu.cn}
\affiliation{$^1$State Key Lab for Mesoscopic Physics, School of Physics,
Peking University, Beijing 100871, P. R. China}
\affiliation{$^2$Key Lab of
Quantum Information, University of Science and Technology of China, Hefei
230026, Anhui, P. R. China}

\begin{abstract}
We propose a kind of plasmonic whispering gallery modes highly
localized on the exterior surface of a metal-coated microresonator. This
exterior (EX) surface mode possesses high quality factors at room
temperature, and can be efficiently excited by a tapered fiber. The EX mode
can couple to an interior (IN) mode and this coupling produces a strong
anti-crossing behavior, which not only allows conversion of IN to EX modes,
but also forms a long-lived anti-symmetric mode. As a potential application, the EX
mode could be used for a biosensor with a sensitivity high
up to $500$ nm per refraction index unit, a large figure of merit, and a wide
detection range.
\end{abstract}

\date{\today }
\maketitle

\preprint{}

Optical whispering-gallery (WG) microresonators provide a
powerful platform for various photonic applications ranging from
low-threshold lasing to highly sensitive bio/chemical sensing \cite{Vahala}.
They are also used for fundamental studies including cavity optomechanics
\cite{Kipp}, cavity quantum electrodynamics (QED) \cite{Aoki,Wang} and
quantum information science \cite{Xiao,Edo} in the past few years.
WG modes are dominantly confined in the high-refraction-index dielectric
material, i.e., the inside of the cavity body. The other few energy of the
WG mode is stored in the weak exterior evanescent field with a
characteristic length of tens to hundreds of nanometers. The evanescent
field is of importance because it offers the effective pathway to exchange
the energy between the cavity mode and the external devices, i.e., the
near-field coupling.

Furthermore, important applications of WG modes lie on
the evanescent field since it determines the light-matter interaction
strength. For instance, the resonant wavelength of the WG mode is sensitive
to the refractive index change induced by the binding of biological/chemical
molecules to the resonator surface. Thus, WG microresonators can be used for
highly sensitive detection of single biological/chemical molecules, which
attracts a strongly increasing attention very recently \cite{Vollmer,Fan}.
Nevertheless, the detection sensitivity is strongly limited by the weak
exterior evanescent field. In this Letter, we propose a new kind of WG modes
of a metal-coated silica microcavity. Unlike the conventional WG modes
distributing inside of the cavity body, the proposed new WG modes are
concentrated in the exterior surface of the coated microcavity. Assisted by
the propagating surface plasmon, these exterior (EX) modes show a high
localization and reasonable high quality factors.

\begin{figure}[tb]
\centerline{\includegraphics[keepaspectratio=true,width=0.4%
\textwidth]{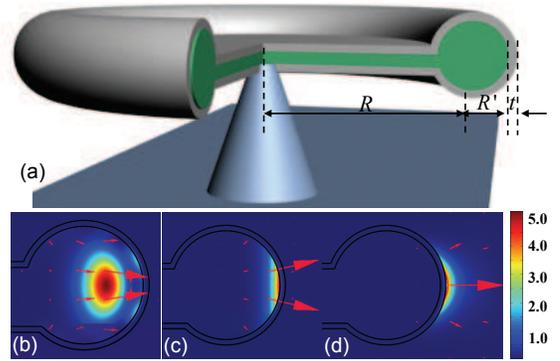}}
\caption{(Color online) (a) Schematic illustration of a metal coated silica
toroidal microcavity supported by a silicon pillar. (b)-(d) False-color representations of the squared
transverse electric fields for optical, IN and EX
plasmonic WG modes, respectively, where red arrows show the directions of
the electric fields. In optical mode, a minor mode hybridization exists.}
\end{figure}

The geometry of the proposed metal-coated silica microcavity is shown in
Fig. 1(a), where $R$ and $R^{\prime }$ are the major and minor radii of the silica
toroid, $t$ represents the metal thickness. The boundary radius is defined as $%
R_{b}=R+R^{\prime }+t$. At room temperature, the permittivities of silica and metal (here we
choose silver as an example) are $\epsilon _{1}=1.4457^2$ and $\epsilon
_{2}=-22.69+0.233i$ in the $680$ \textrm{nm} band,
respectively. The permittivity of the surrounding dielectric material is
denoted as $\epsilon _{3}=n_{3}^{2}$. In the following study, on one hand, we
fix $R^{\prime }=1$ $\mathrm{\mu m}$, $\epsilon _{1}$ and $\epsilon _{2}$,
but vary the parameters $R$, $t$, and $\epsilon _{3}$. On the other hand,
since the EX mode has the great potential in biosensing, $n_{3}$ is set
around $1.33$ (the refraction index of water). To intuitively exhibit the EX
modes, here we first resort to a full-vectorial eigenmode solver \cite%
{Oxborrow}. Figures 1b-1d show field intensity distributions $\left\vert E(%
\vec{r})\right\vert ^{2}$ of three typical WG modes. The first is inside of
silica, which is the conventional optical WG mode. The second is localized
in the interior surface of the coating, and the similar interior (IN)
plasmonic modes have been investigated in a metal-coated silica microdisk
recently \cite{Bumki,Painter}. The last is localized in the exterior surface
of the coating, which is the focused scope of the present paper. In general,
the EX modes can be tentatively identified by two parameters: the angular
mode number $l$ and the azimuthal mode number $m$. Due to the
surface-plasmon property, they are inherently quasi-TM and fundamental
radial modes. In this Letter, we focus on the fundamental azimuthal modes,
i.e., $m=l$.

\begin{figure}[phtb]
\centerline{\includegraphics[keepaspectratio=true,width=0.45%
\textwidth]{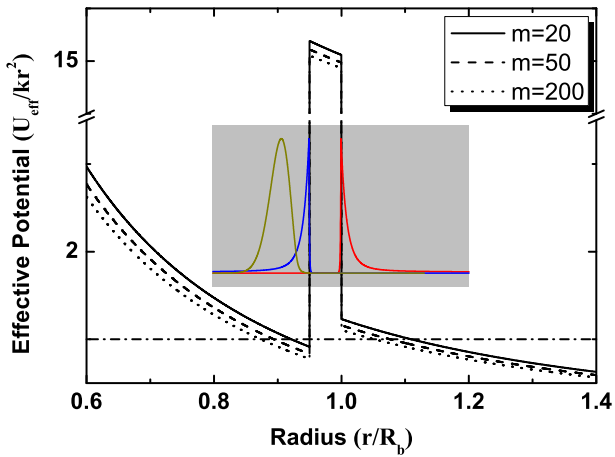}}
\caption{(Color online) Effective potential function $U_{\mathrm{eff}%
}(r)/kr^{2}$ for a layered 2D resonator with a silica core coated by a
negative silver layer, where $k=2\protect\pi /\protect\lambda $. The
horizontal dash-dot line shows the normalized photon energy. The radial
distributions of optical mode, IN and EX plasmonic modes are shown in
yellow, blue, and red curves, respectively. Here $n_{3}=1.33$.}
\end{figure}

Before further investigating the EX mode numerically, we turn to
theoretically analyze the confinement mechanism. Here a 2D model is adopted
for simplification because there is no analytical solution to the toroid
shaped resonator and the 2D model could be a good approximation as the EX
mode is on the equatorial plane. It is obvious that both EX and IN modes are
plasmonic modes which propagate along the metal-dielectric interface. Their
energies are stored in the form of the collective oscillation of electrons
in metal and evanescent wave in the surrounding dielectric. The effective
potential approach \cite{Potential} provides a good physical insight into
many properties of the WG modes that appear as quasibound states of light,
analogous to the circular Rydberg states of alkali atoms. Thus, in Fig. 2,
we plot the effective potential of an EX mode of a layered 2D resonator
(very similar results can be obtained for the IN plasmonic and conventional
optical WG modes). It can be found that the negative permittivity of metal
leads to a potential barrier inside the cavity. This configuration of the
effective potential produces three kinds of WG modes: conventional optical
mode, IN and EX plasmonic modes, corresponding to Figs. 1b, 1c and 1d.

\begin{figure}[phtb]
\centerline{\includegraphics[keepaspectratio=true,width=0.45%
\textwidth]{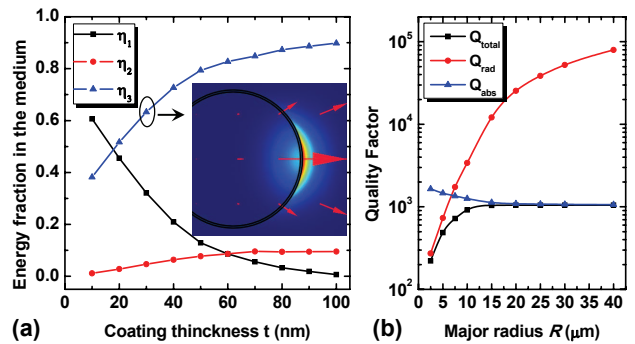}}
\caption{(Color online) (a) Mode energy distribution fractions of an EX WG
mode $\protect\eta _{1,2,3}$ in silica, metal and surrounding medium,
respectively. Here $R=10$ $\mathrm{\protect\mu m}$ and $n_{3}=1.33$. Inset:
The mode distribution when $t=30$ \textrm{nm}. (b) Total, radiation-related,
and absorption-related quality factors, $Q_{\mathrm{total}}$, $Q_{\mathrm{rad%
}}$, and $Q_{\mathrm{abs}}$, depending on the major radius $R$. Here $t=100$
\textrm{nm} and $n_{3}=1.33$.}
\end{figure}

To characterize EX modes in more detail, we now investigate the spatial
distributions through the full-vectorial eigenmode solver. The mode energy
fractions in the inner silica, metal nanolayer, and outer surrounding are
denoted as $\eta _{1,2,3}$, respectively. By changing the metal thickness $t$,
the curves in Fig. 3a illustrate the following points. (i) The EX mode has
considerable distributions in both the inner silica and outer medium in the
case of a thin coating, for example, $\eta _{1}=0.60644$ and $\eta
_{3}=0.38193$ when $t=10$ \textrm{nm}; while the minor energy is in the
coating nanolayer ($\eta _{2}=0.01163$). This phenomenon reflects the fact
that the EX mode can tunnel to the inner through the potential barrier shown
in Fig. 2. (ii) The fraction $\eta _{1}$ decreases quickly with increasing
the coating thickness. In particular, $\eta _{1}$ approaches $0$ for a large
enough thickness (for example, $t=100$ \textrm{nm}). (iii) As expected, more
and more energy distributes in the exterior of the microcavity when the
coating thickness increases. For example, $\eta _{3}$ reaches $90\%$ at $%
t=100$ \textrm{nm}. Qualitatively, the inset picture in Fig. 3a, with $t=30$
\textrm{nm}, shows that the EX mode has invaded the interior silica. In
addition, the EX mode is highly localized on the exterior surface of the
coated cavity. The effective mode area is defined as $A_{\mathrm{eff}}\equiv
\int_{A}u(\vec{r})dA/u(\vec{r})|_{\max }$, where $u(\vec{r})=\left[ d\left(
\omega \epsilon (\vec{r})\right) /d\omega \left\vert E(\vec{r})\right\vert
^{2}+\mu _{0}\left\vert H(\vec{r})\right\vert ^{2}\right] /2$ gives the mode
energy density, with the electric and magnetic fields $E(\vec{r})$ and $H(%
\vec{r})$, respectively, the dielectric permittivity $\epsilon (\vec{r})$
and the permeability of vacuum $\mu _{0}$. $A_{\mathrm{eff}}$ is typically
smaller than $0.1$ $\mathrm{\mu m}^{2}$ with the coating thickness $t$ ranging
from $10$ to $100$ \textrm{nm}, much smaller than the conventional optical
WG mode with the same cavity size.

Quality factor of the mode, associated with the photon lifetime in the cavity,
is one of the important parameters of an optical resonator. Here we not only
provide the total quality factor $Q_{\mathrm{total}}$, but also separately
calculate the radiation and metal absorption related quality factors $Q_{%
\mathrm{rad}}$ and $Q_{\mathrm{abs}}$, respectively, though these two
contributions can not be separated experimentally. Thus it allows
systemically studying of the different dissipation mechanisms of the EX mode.
Here the loss caused by silica or water absorption has been omitted since it
is much smaller than the silver absorption. As shown in Fig. 3b, $Q_{\mathrm{%
rad}}$ (by assuming pure real silver permittivity) increases while $Q_{%
\mathrm{abs}}$ decreases with increasing the cavity major radius $R$. For $%
Q_{\mathrm{rad}}$, it is reasonable since the large size will reduce the
radiation loss; while for $Q_{\mathrm{abs}}$, its decrease results from that
more and more energy distributes in the metal layer. These two results can
also be understood from Fig. 2. With increasing the cavity size (thus
increasing the mode number $m$), the potential barrier is lowering, and the
light has a larger penetration depth in the metal. Thus more energy will be
confined in the metal, and this suppresses $Q_{\mathrm{abs}}$. Moreover, the
increasing energy in the metal results in a larger momentum of light. This
makes the light more difficult to escape or couple to free space due to the
momentum mismatching (i.e., reduce the radiation loss). It is easy to find
that at a very large $R$, the total quality factor is limited by the metal
absorption loss, i.e., $Q_{\mathrm{total}}\sim Q_{\mathrm{abs}}$; while for
a small $R$, $Q_{\mathrm{total}}$ mainly depends on the radiation loss. The
total quality factor reaches $1000$ when the major radius is $10$ $\mathrm{%
\mu m}$, which is much larger than most plasmonic microcavities \cite%
{spp1,spp2,spp3,spp4,spp5}.

\begin{figure}[phtb]
\centerline{\includegraphics[keepaspectratio=true,width=0.45%
\textwidth]{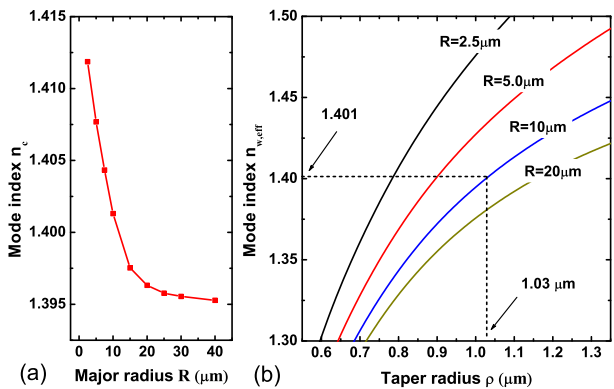}}
\caption{(Color online) (a) EX WG mode index $n_{c}$ for the different major
radius. (b) Transformed mode index $n_{w,tr}$ of the straight fiber taper
waveguide in the curved geometry of the cavity. Curves from left to right
correspond to $R=2.5,5,10,20$ $\mathrm{\protect\mu m}$, where the taper is
attached on the surface the coated microcavity. Here $t=100$ \textrm{nm} and
$n_{3}=1.33$.}
\end{figure}

The energy exchange (i.e., coupling) between an optical microcavity and an
external device is of importance since it directly affects most
applications. To realize an efficient coupling, there are two essential
requirements: (i) sufficient spatial overlap of the cavity mode field and
external device, (ii) phase matching condition between them. Here we propose that
a tapered fiber waveguide with waist diameter $2\rho $ can implement such a
near-field coupling \cite{cai}. For the requirement (i), the EX WG mode
possesses an inherent advantage because the mode field dominantly
distributes in the exterior of the cavity body. As a result, it allows
building a very strong spatial overlap between the EX and waveguide modes.
Here we consider that the waveguide is attached to the cavity to obtain the
maximum coupling strength. For the requirement (ii), it is necessary to
compare the cavity and waveguide mode indexes $n_{c}$ and $n_{w}$. The
cavity mode index $n_{c}$ of a specific eigenmode (angular mode number $l$,
resonant wavelength $\lambda $) can be evaluated with respect to the cavity
edge as $n_{c}=l\lambda /(2\pi R_{b})$. As shown in Fig. 4a, $n_{c}$
gradually decreases from $1.41187$ to $1.39527$ with increasing the cavity
major radius $R$. This decrease is due to the choice of the cavity edge $%
R_{b}$. A rigorous calculation should consider the effective radius $R_{e}$,
which is typically larger than $R_{b}$ for EX modes. Nevertheless, the
present $n_{c}$ is still valid because both $n_{c}$ and $n_{w}$ are
evaluated based on $R_{b}$.

It should be emphasized here that $n_{c}=n_{w}$ is not the exact phase
matching condition because of the curved and straight geometries of the
cavity and waveguide. As seen by the curved cavity, $n_{w}$ has an effective
mode index $n_{w,\mathrm{eff}}=n_{w}/\left( 1+\delta /3+2\delta
^{2}/15+O(\delta ^{3})\right) $ with $\delta =-\rho /R_{b}$ \cite{Rowland}.
Thus, the exact phase matching condition should be $n_{c}=n_{w,\mathrm{eff}}$%
. Figure. 4b displays the effective waveguide mode index $n_{w,eff}$
depending on $R$ and $\rho $. A thicker tapered waveguide will result in a
larger mode index $n_{w,eff}$. As we can see from these curves, $n_{w,%
\mathrm{eff}}$ easily covers $1.3$ to $1.5$ for a micron-sized taper. For
instance, when $R=10$ $\mathrm{\mu m}$, $n_{c}$ is about $1.401$. To achieve
the phase matching condition, the taper radius can be around $1.03$ $\mathrm{%
\mu m}$, which is attainable with current technique.

\begin{figure}[phtb]
\centerline{\includegraphics[keepaspectratio=true,width=0.45%
\textwidth]{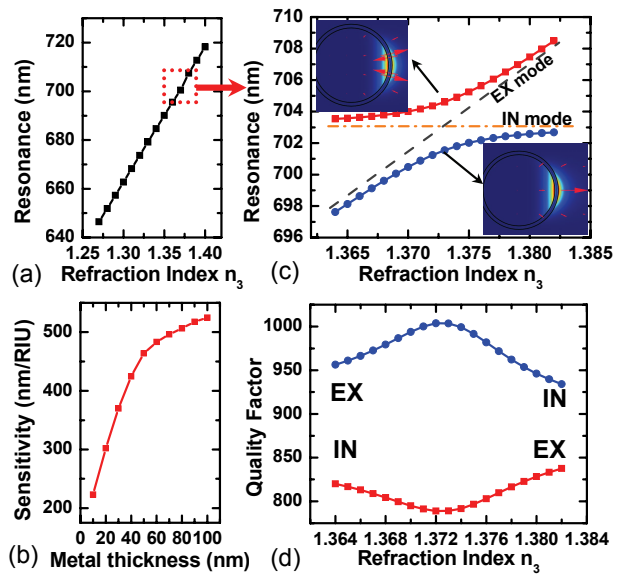}}
\caption{(Color online) (a) Resonant wavelength of an EX mode vs. the
refraction index $n_{3}$. Here $R=10$ $\mathrm{\protect\mu m}$
and $t=100$ nm. (b) The sensitivity depending on the coating thickness. (c)
Anti-crossing phenomenon of the resonant wavelengths when $n_{3}$ is around $%
1.37$. As a guide the eye, black dash and orange dash-dot lines show the
wavelengths of the uncoupled EX and IN WG modes, respectively. (d) The
quality factors of the modes corresponding to (c). The curves are fitted with Eq. (1).}
\end{figure}

As mentioned above, the EX mode may be applied in highly sensitive
biosensors. To demonstrate this potential, Fig. 5a plots the resonant
wavelength of a specific EX mode for different refraction index $n_{3}$ of
the surrounding dielectric (this can simulate the binding of
biological/chemical molecules to the resonator surface). It shows a highly
linear increase with increasing $n_{3}$. By calculating the slope, we can
obtain the sensitivity of the biosensing, defined as $S=d\lambda /dn_{3}$.
Figure 5b shows that the sensitivity monotonously increases with the coating
thickness. This result can be understood by considering the mode energy
fraction $\eta _{3}$ in the surrounding dielectric (blue line in Fig. 3a).
It can be found that the sensitivity of the EX mode-based sensor exceeds $%
500 $ \textrm{nm/RIU} (Refraction index unit).
Furthermore, the overall figure of merit (FoM) defined by the ratio of the
sensitivity and the resonance linewidth, is also studied. Since the EX mode
has a high quality factor of $1000$, the corresponding FoM reaches $700$,
exceeding (at least highly comparable to) the propagating surface plasmon
\cite{homola} or the fiber Bragg grating \cite{chry}\ based biosensing.

It is interesting to note that a minor nonlinearity occurs when $n_{3}$ is
around $1.37$. To explicitly expose the underlying physics of the WG mode in
this area, we further calculate the resonant wavelengths of the modes with a
much smaller $n_{3}$ change ($0.001$). As shown in Fig. 5c, a clear
anticrossing occurs where two energy eigenvalues come near to
cross but then repel each other, similar to the strong interaction
between a microcavity resonance mode and a two-level quantum dot. This
is the signature of a strong coupling between the fundamental EX and IN WG
modes, as depicted by the mode distribution in the insets of Fig. 5c. The
strongest coupling occurs at $n_{3}\approx 1.373$ where not only the two
modes are spatially overlapped in the metal nanolayer but also their mode
indexes are exactly matched. On the left side of this point, the red data
points are the IN-like modes while the blue points represent the EX-like
modes, and it is opposite on the right side. Remarkably, this anticrossing
behavior, arising from the phase-matched interaction between the EX and IN
modes, offers an efficient pathway to convert the IN mode to the EX mode.
Furthermore, the quality factors of these two coupled modes are plotted in
Fig. 5d. It shows a strong modulation by the coupling and forms a long-lived
anti-symmetric mode (upper curve) due to the external coupling \cite{cao}.

In general, this mode coupling can be described in terms of a $2\times 2$
Hamiltonian matrix with eigen energies of the two modes $E_{1}$ and $E_{2}$, and
the coupling terms $V_{1,2}$. Under the coupling, it forms two new eigen states
(symmetric and asymmetric modes) with energies $E=\left( E_{1}+E_{2}\right)
/2\pm \sqrt{\left( E_{1}-E_{2}\right) ^{2}/4+V_{1}V_{2}}$. Fitting with this
theoretical description, we obtain the coupling strength $\sqrt{\left\vert
V_{1}V_{2}\right\vert }\sim 1$ \textrm{THz} ($3.8$ \textrm{meV}).

In summary, a new kind of WG modes is theoretically studied to be highly
localized on the exterior surface of a metal-coated silica toroid microcavity.
The EX WG modes possess a high quality factor up to one thousand at room temperature
and can be efficiently excited by a tapered fiber. The coupling
between IN and EX modes is also investigated. It is found that the coupling
not only produces a strong anti-crossing, but also forms a long-lived anti-symmetric mode.
Potentially, these EX modes are a widely useful advance beyond the IN modes described in Ref. \cite%
{Bumki}, for instance, highly sensitive biosensors with a wide detection range.

The authors acknowledge support from NSFC (Nos. 10821062 and 11004003), the National Basic Research
Program of China (Nos. 2006CB921601 and 2007CB307001). YFX was also supported by RFDP (No. 20090001120004) and SRF for ROCS, SEM.

\end{document}